\documentclass[]{aa}
\def\etal{{et\,al. }}
\usepackage{graphicx}
\begin{document}
   \title{INTEGRAL results on GRB030320: a long gamma-ray burst detected at the edge of the field of 
view\thanks{Based on observations with INTEGRAL, an ESA project
with instruments and science data centre funded by ESA
member states (especially the PI countries: Denmark, France,
Germany, Italy, Switzerland, Spain), Czech Republic and
Poland, and with the participation of Russia and the USA and on observations made with ESO telescopes at the La Silla
Observatory under programme Id 71.D-0667(C)}}

   \author{
 A. von Kienlin\inst{1} \and
 V. Beckmann\inst{2,3} \and
 S. Covino\inst{8}\and
 D. G\"{o}tz \inst{4,5} \and
 G. G. Lichti \inst{1} \and
 D. Malesani \inst{9}\and
 S. Mereghetti \inst{4} \and
 E.~Molinari \inst{8}\and
 A. Rau \inst{1} \and
 C. R. Shrader \inst{6,7}\and
 S. J. Sturner \inst{6,7} \and
 F. Zerbi \inst{8}
}
\authorrunning{A. von Kienlin et al.}
\titlerunning{GRB030320: A long GRB at the edge of INTEGRAL's field of view}

   \offprints{A. von Kienlin \\ (\email{azk@mpe.mpg.de}) }

   \institute{ 
Max-Planck-Institut f\"{u}r extraterrestrische Physik, Giessenbachstrasse, 85748 Garching, Germany \and
%
%
Institut f\"ur Astronomie und Astrophysik, Universit\"at T\"ubingen, Sand 1, D-72076 T\"ubingen, Germany \and
INTEGRAL Science Data Centre, Chemin d' \'Ecogia 16, CH-1290 Versoix, Switzerland \and
Istituto di Astrofisica Spaziale e Fisica Cosmica - CNR, Sezione di Milano ``G.Occhialini'', Via Bassini 15, I-20133 Milano, Italy \and
Dipartimento di Fisica, Universit\`{a} degli Studi di Milano Bicocca, P.zza della Scienza 3, I-20126 Milano, Italy \and
Code 661, NASA/Goddard Space Flight Center, Greenbelt, MD 20771, USA \and
Universities Space Research Association, 7501 Forbes Blvd. \#206, Seabrook, MD
20706, USA \and
INAF / Brera Astronomical Observatory, Via E. Bianchi 23807, Merate (LC),
Italy \and
International School for Advanced Studies (SISSA-ISAS), Via Beirut 2-4,
34014 Trieste, Italy
} 
  
   \date{Received --; accepted --}

   \abstract{GRB030320 is the 5th Gamma-ray burst (GRB) detected by INTEGRAL in the field of view (FoV). 
It is so far the GRB with the largest off-axis angle with respect to the INTEGRAL pointing direction, 
near to the edge of the FoV of both main instruments, IBIS and
SPI. Nevertheless, it was possible to determine its position and to extract spectra and fluxes. 
The GRB nature of the event was confirmed by an IPN triangulation. 
It is a $\sim$\,60 s long GRB with two prominent peaks
separated by $\sim$\,35 s. The spectral shape of the GRB is best
represented by a single power law with a photon index $\Gamma \simeq 1.7$. The
peak flux in the 20 -- 200 keV band  is determined to $\sim 5.7$ photons
cm$^{-2}$ s$^{-1}$  
and the GRB fluence to  $ 1.1 \times 10^{-5}$ erg cm$^{-2}$. 
Analysing the spectral evolution of the GRB, a ``hard-to-soft'' behaviour emerges. 
A search for an optical counterpart has been carried out, but none was found.

   }
                      
   \maketitle
%

\section{Introduction}   

\label{sect:intro}  

Since their discovery more than thirty years ago (\cite{Klebesadel73}) GRBs
had evaded an understanding of their nature for about 20 years. This
changed dramatically with the identification of the first GRB X-ray
afterglow with a BeppoSAX observation (\cite{Costa97}). Since then it has
been shown that at least the sources of long GRBs (lasting longer than 2 s) are 
located at cosmological distances. The observed fluxes revealed the enormous 
energy release of these events (\cite{paradijs2000}) which is thought to be produced in 
asymmetric collapse of massive stars (\cite{woosley93}). 

Here we report the detection of such an event by the two main
instruments of INTEGRAL (\cite{integral}) which have complementary
performance characteristics. The imager IBIS (\cite{ibis}) has an
excellent angular resolution, whereas the spectrometer SPI (\cite{spi})
is assigned to high-resolution spectroscopy, but has only modest
imaging capabilities. The burst-detection capabilities of the INTEGRAL
mission are achieved by the INTEGRAL burst alert system (IBAS: \cite{ibas}, IBAS for SPI/ACS: \cite{Kienlin01}),
which scans the satellite telemetry in near-real time for GRBs at the
INTEGRAL Science Data Centre (ISDC; \cite{isdc}).

  \begin{figure}
   \centering
   \includegraphics[width=0.45\textwidth,bb= 50 150 550 670,angle=270]{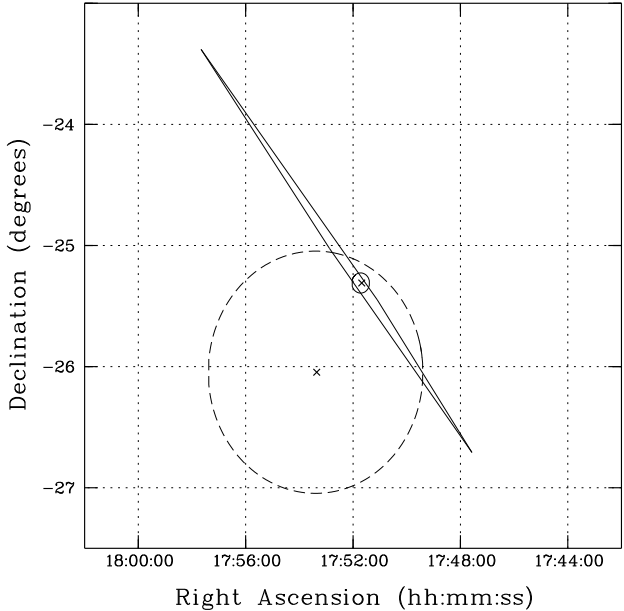} 
   \caption {
      Localisations of GRB030320 showing the IBIS/ISGRI position (central cross) and error circle (solid line) 
of Mereghetti et al. (2003b). The IPN error box of Hurley et al. (2003b) and the derived SPI position with its $1^{\circ}$
error circle (dashed line) are superimposed.}
         \label{fig:GRB-loc} 
   \end{figure}

\section{Detection and Localisation}

On March 20, 2003 at 10:11:40 UTC, during the Galactic Centre Deep
Exposure, a long GRB ($\sim 60$ s) was detected at a large off-axis angle
of $15.5^\circ$. This is near to one corner of the squared shaped IBIS FoV: $29^\circ \times 29^\circ$ full 
width (SPI-FoV: $\sim 35^\circ$ full width). 

The GRB was detected by IBAS in near-real time in the data of ISGRI (\cite{isgri}), the 
low-energy detector of IBIS. As the trigger had a low
significance due to the position of the source on the edge of the FoV (only 3.7\% of
the detector was illuminated by the GRB) no prompt alert was
distributed. An offline interactive analysis of the data confirmed
that the trigger was due to an astrophysical source and allowed to determine and
distribute a preliminary position 6.25 hours after the event (IBAS Alert
343; \cite{gcn1941}). The  preliminary  derived position was 
$\alpha_{\rm J2000} =$ $17^{\rm h}$~$51^{\rm m}$~$43^{\rm s}$, $\delta_{\rm J2000} =$~$-25^{\circ}$~$18'$~$34''$
with an uncertainty of $5'$. 
In a further analysis of the entire GRB (UTC 10:11:36 - 10:12:40) in the 15-100 keV energy band 
the GRB was detected with a signal-to-noise ratio of S/N $\sim15$. The derived error box had nearly the same centre 
$\alpha_{\rm J2000} =$ $17^{\rm h}$~$51^{\rm m}$~$42^{\rm s}$, $\delta_{\rm J2000} =$~$-25^{\circ}$~$18'$~$44''$ but a 
smaller radius of $3'$.

The GRB nature of the event was confirmed about one hour later by Konus, Mars Odyssey (HEND), and INTEGRAL/SPI-ACS via 
an IPN-annulus 
(\cite{gcn1942}). By adding the information from Ulysses, Mars Odyssey (GRS), Konus-Wind and RHESSI to the above mentioned  
data, 
a small IPN error box was derived one day later (\cite{gcn1943}).

SPI was only able to determine a rough position for the GRB in the energy
interval between $100\, \rm keV$ and 1 MeV. At lower energies it was not
possible to derive a position. One should notice  that only 3 out of the 19
detectors of SPI's camera showed a substantial increase of the count rate,
caused by the illumination of the GRB. 
By selecting two time intervals around the prominent peaks of emission (see below) from UTC 10:11:55 to 10:12:05 and  
10:12:30 to 10:12:36 (in total 16\,s), the derived SPI position is 
$\alpha_{\rm J2000} =$ $17^{\rm h}$~$53.4^{\rm m}$, $\delta_{\rm J2000} =$~$-26^{\circ}$~$2.8'$ with an uncertainty 
of $1^{\circ}$ (S/N 7.2). This agrees within the error with the 
positions found by IBIS/ISGRI and IPN.
Fig.~\ref{fig:GRB-loc} shows a superposition  of the IPN error box and the SPI- and IBIS/ISGRI-error circles.

The two monitoring instruments of INTEGRAL, the X-ray monitor JEM-X (\cite{jemx})
and the optical camera OMC (\cite{omc}), did not observe the event as it was outside their FoVs.

A search for an optical afterglow in the IBIS error circle was performed
using the Wise-Observatory 1 m telescope 16.9 hours after the onset of the
burst (\cite{gcn1946}). This follow-up observation did not reveal
any optical counterpart. R-band imaging under poor conditions did not
show any new source down to a limiting magnitude of $R = 17.5 \, \rm
mag$.

%
  \begin{figure}
   \centering
   \includegraphics[width=0.5\textwidth,angle=270]{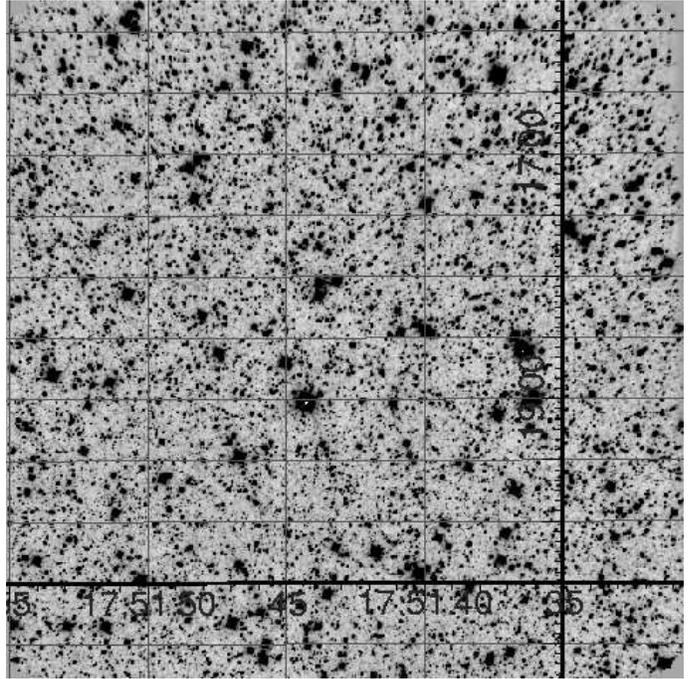} 
   \caption {
ESO/NTT+SOFI image corresponding to the centre of the IBIS error circle
derived here. The field is 5.5$\times$5.5 arcmin wide.
The 5 $\sigma$ limiting magnitude is {\it Ks}=20.7.
}
         \label{fig:ntt-obs} 
   \end{figure}

We also performed two follow-up observations at ESO New Technology
Telescope (NTT) using the SOFI camera.
The first was made at 9.497 UT on the 21$^{\rm st}$ of March
($\sim$ 23.3 hours after the GRB) with a {\it Ks} filter and had exposure
of 30\,min (see Fig. \ref{fig:ntt-obs}).
The second was taken at 4.608 UT ($\sim$ 69.8\,days after the GRB)
on the 29$^{\rm th}$ of May with the same filter and an exposure of 10\,min.
The difference image, although showing several variable sources,
did not reveal a strong afterglow candidate.

The afterglow search was complicated by the burst location in the galactic
plane. The foreground galactic hydrogen column 
density\footnote{http://heasarc.gsfc.nasa.gov/cgi-bin/Tools/w3nh/w3nh.pl}
of $N_{\rm H}$=(1.4$\pm$0.1)$\times$10$^{22}$\,atoms/cm$^2$
in the direction of the burst results in an extinction of  
$A_{\rm V} = (10.5 \pm 0.5) \, \rm mag$ [$A_{\rm R} = (7.85 \pm 0.35) \, \rm mag$].
This high extinction would have required an extraordinary intrinsically
bright afterglow ($R_{\rm int} < 10 \, \rm mag$ at the time of the Wise
observations) in order to be detectable. This would be more than 5\,mag
brighter than the very bright afterglow of GRB030329 (\cite{grb030329}) at the
same time after the burst occurrence.

\section{Temporal and Spectral Characteristics} 

  \begin{figure}
   \centering
   \includegraphics[width=0.5\textwidth,bb= 50 65 540 770,angle=0]{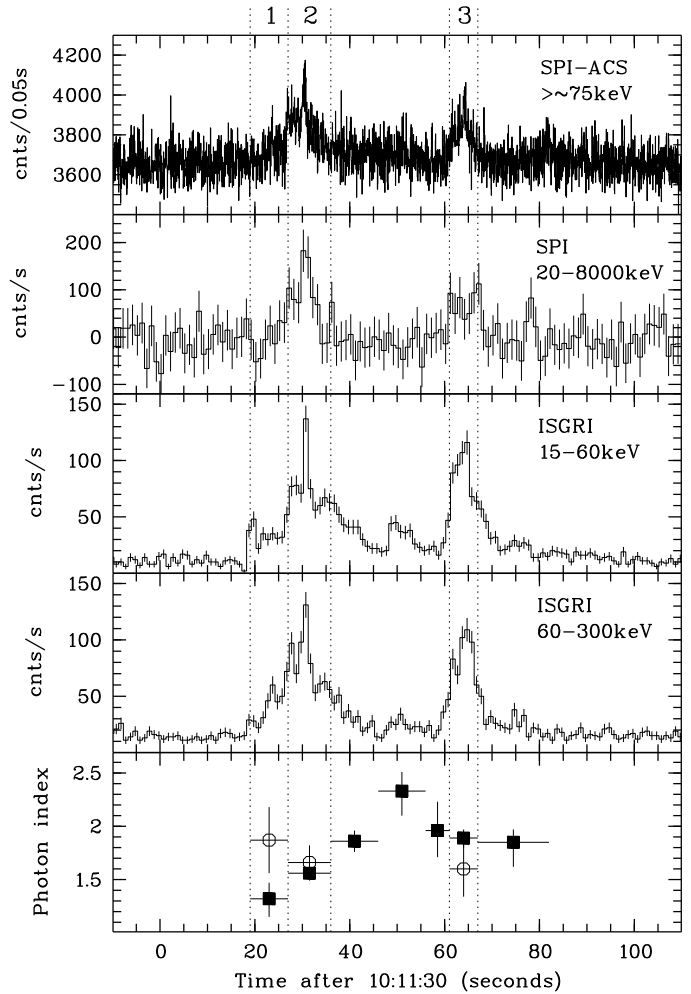} 
   \caption {Lightcurves of all INTEGRAL instruments which have seen GRB030320, from top to bottom: SPI-ACS 
overall veto count rate with 50\,ms time resolution, SPI 1\,s binned and IBIS/ISGRI in a soft (15-60 keV) and a hard 
(60-300 keV) energy range with 1\,s binning. The bottom plot shows the evolution of photon index 
$\Gamma$ [$f(E) = K \cdot (\frac{E}{1 {\rm keV}}^{-\Gamma})$;
$ K= $ photons/(keV cm$^2$ s) at 1 keV] for ISGRI  (filled square) and SPI  (XSPEC 12.0 results, open circle). 
The dotted lines mark the intervals 1 -- 3 from Tab. \ref{tab:specflux}.}
         \label{fig:spi-ibis-lc} 
   \end{figure}

\begin{table*} 
\centering
\caption[]{Photon indices and fluxes obtained by IBIS/ISGRI and SPI (by three different methods, see text) 
in 6 time intervals of the GRB. The quoted uncertainties are given for a 90\% confidence range. For line SPI$_{12.0}$ it was only possible 
to quote the 90\% confidence range approximately, because the calculation of uncertainty contours is not yet implemented 
in XSPEC 12.0. 
}
\label{tab:specflux}
\begin{small}
\begin{tabular}{|c|l||c|c|c||c|c|c|} 
\hline 
\multicolumn{2}{|c||}{\rule[-1ex]{0pt}{3.5ex}Interval \# }  & 1 & 2 & 3 & 4 & 5 & 6\\ \hline
\multicolumn{2}{|c||}{\rule[-1ex]{0pt}{3.5ex} Position }  & before $1^{\rm st}$~peak &
$1^{\rm st}$ peak & $2^{\rm nd}$ peak & max $1^{\rm st}$ peak &  \#1 $+$ \#2 & full GRB \\ \hline
\multicolumn{2}{|c||}{\rule[-1ex]{0pt}{3.5ex}UTC 10 : \_\_ : \_\_ }  &  11:49-11:57 &
11:57-12:06 &  12:31-12:37  &  12:00-12:02 & 11:49-12:06  &  11:49-12:37\\
\hline \hline
 \rule[-1.5ex]{0pt}{4ex} & ISGRI & 1.32 ${0.15 \atop -0.17}$ & 1.56 ${0.08 \atop -0.07}$ &  1.89 ${0.09 \atop -0.09}$ & 1.62 ${0.08 \atop -0.09}$ &  1.52 ${0.08 \atop -0.08}$ & 1.69 ${0.07 \atop -0.08}$ \\ 
 \rule[-1.5ex]{0pt}{4ex}  & SPI$_{12.0}$ &  $1.87 \pm 0.31$ & $1.66 \pm 0.16$ & $1.60\pm 0.26$ & $1.01 \pm 0.31$  & $1.54 \pm 0.15$ & $1.51 \pm 0.16$ \\ 
  \rule[-1.5ex]{0pt}{4ex} \raisebox{1.5ex}[-1.5ex]{PhoIndex}  & SPI$_{11.2}^{\rm offd.}$ &  1.68 ${0.67 \atop -0.67}$  & 0.94  ${0.45 \atop -0.87}$  & 1.29  ${0.45 \atop -0.64}$ &  $-$ & 1.28  ${0.35 \atop -0.48}$ & 1.19   ${0.29 \atop -0.37}$ \\ 
 \rule[-1.5ex]{0pt}{4ex}  & SPI$_{11.2}^{\rm diag.}$ &  1.97 ${0.71 \atop -0.47}$ &  1.40  ${0.25 \atop -0.28}$ &  1.56  ${0.36 \atop -0.38}$  &  1.03  ${0.43 \atop -0.50}$ &  1.54  ${0.24 \atop -0.25}$ & 1.57  ${0.20 \atop -0.20}$ \\ 
\hline \hline
 \rule[-1.0ex]{0pt}{3.5ex}  20$-$200 keV & ISGRI & 1.449 & 3.841 & 3.685 & 5.689 & 2.773 &
 2.068 \\ 
 \rule[-1.0ex]{0pt}{3.5ex} Flux & SPI$_{12.0}$ & 3.08  &  4.35  &  3.53 & 3.14 & 3.24  &  2.11 \\ 
 \rule[-1.0ex]{0pt}{3.5ex}  & SPI$_{11.2}^{\rm offd.}$ &  3.62 &  2.05  &  3.56 & $-$ &  2.78 &  1.65 \\ 
 \rule[-1.0ex]{0pt}{3.5ex}  \raisebox{1.5ex}[-1.5ex]{$\left[ \frac{\rm ph}{\rm cm^{2} s}\right]$} & SPI$_{11.2}^{\rm diag.}$ &  4.40 & 4.45 &   4.95 &  4.13  & 4.11  & 2.82\\ \cline{2-8}
\hline 
\end{tabular}
\end{small}
\end{table*}

Fig.\,\ref{fig:spi-ibis-lc} shows the lightcurve for all INTEGRAL detectors which have observed GRB030320. 
The lightcurves of ISGRI have been determined in two energy bands (15-60\,keV and 60-300\,keV) 
by using only the pixels that were illuminated by the GRB by at least half of their surface. The 
determination of a SPI lightcurve with a short  time binning was only possible by using the 1\,s 
count rates of the 19 Ge-detectors, which are normally used for science-housekeeping purposes. 
These values reflect the count rates of each detector in the broad SPI energy band from $\sim$\,20\,keV up to $\sim$\,8\,MeV.
The SPI lightcurve in Fig.~\ref{fig:spi-ibis-lc} was generated by summing the background-subtracted 
count rates. The background ($\sim$\,50\,counts\,s$^{-1}$\,detector$^{-1}$) was determined from the SPI
data before the burst occurrence over a duration of 40 min. The alternative
method, which uses the time-tagged photon-by-photon SPI mode, with energy
information for each photon, yielded insignificant results.
With the anti-coincidence shield (ACS) of SPI the GRB was
observed as an increase of the overall veto count rate. 
The effective area of the ACS is very small for sources observed 
within the field of view (\cite{spie2003}).
For GRBs which occur outside the FoV of SPI the effective area of the ACS is however large, 
as it can be seen from the detection rate of
$\sim$\,0.8 GRBs/day (\cite{spi-grbs}).
All lightcurves in Fig.\,\ref{fig:spi-ibis-lc} exhibit two prominent peaks during the $\sim$\,60\,s of 
prompt emission. A third peak is visible in the low-energy
range of ISGRI, which is less significant at higher energies.

Spectra were extracted for characteristic intervals of the GRB lightcurve. 
The GRB spectrum was in all cases well represented by a single power-law model. The
photon index and the source flux in the 20--200 keV energy range are
listed for ISGRI and SPI in Tab.~\ref{tab:specflux}.
For both instruments, the response for a source position near-to-zero coding is not yet well understood. 
For ISGRI the closest Crab observation was used for the calibration of the spectra. 
Any deficiencies of the SPI response at $15.5^\circ$ off-axis angle is possibly the result of a not 
fully-representative mass model.
Currently the off-axis effects of the response are assessed by using Crab observations, recorded during 
the payload-verification campaign, 
but the drawback of this method is the lack of Crab observations at such large off-axis angles. 

The photon indices and fluxes listed in Tab.\,\ref{tab:specflux} for SPI were obtained by binning the photon-by-photon 
single and multiple events into six equally-spaced logarithmic energy bins in the 20\,keV to 2\,MeV range, 
for each of the listed 
time intervals. 
Spectral extraction was performed using SPIROS (SPI Iterative Removal of
Sources; Skinner \& Connell 2003), which takes the SPI photopeak effective
area into account. Spectral model fitting was performed using XSPEC 11.2
and the off-diagonal response of SPI (SPI-RMF), as shown in the line SPI$_{11.2}^{\rm offd.}$
of Tab.~\ref{tab:specflux}. 
For the background, all event data of the corresponding science window, 
with a duration of about 30\,min, were used, but with the time of the GRB cut out (UTC 10:11:40 - 10:14:00).
Tab.\,\ref{tab:specflux} lists the obtained photon indices and
fluxes, when using XSPEC 11.2 only with a diagonal response (line SPI$_{11.2}^{\rm diag.}$) too.
Naturally
one obtains a softer spectral shape with this method. 

For comparison, the data were analysed using an alpha-test version of the XSPEC 12.0 software. 
This approach is distinct from the SPIROS method, in that the full diagonal plus off-diagonal response matrix 
is computed for each detector (and for each pointing direction, although here the GRB is contained within a single pointing) 
(\cite{sturner2003}; also \cite{shrader2000}).  
The detector-count spectra are then compared directly to the convolution of a photon model with, in this case, 
19 response matrices, 
and a $\chi^2$ minimisation is performed.  This is, in principle, a more rigorous deconvolution procedure 
than the SPIROS method. 
However, the disadvantage is that the source and background  must be simultaneously modelled whereas SPIROS reduces the problem 
to a single background subtracted spectrum and response matrix. 
A set of ancillary response functions (ARFs), containing the energy, detector and angle-dependent effective area, 
were constructed 
for the GRB position and the burst spectrum was modelled in XSPEC 12.0 by fixing the background 
(derived by using the non-burst 
part of the observation). The obtained photon indices and fluxes are listed in the line SPI$_{12.0}$ of Tab.~\ref{tab:specflux}.

ISGRI single events have been used to derive photon indices and fluxes shown in the line ``ISGRI'' of Tab.~\ref{tab:specflux}. 
Thanks to the coded-mask design of the IBIS telescope, the source flux can be determined simultaneously
with the background using the Pixel Illumination Function (PIF; \cite{skinner95}). 
The spectra have been extracted in 128 linearly-spaced energy bins
between 19 keV and 1 MeV and were rebinned to have a signal-to-noise ratio
larger than 3 in each channel. The photon spectra have then been obtained
with the technique described above.
In addition to the time intervals shown in Tab.\,\ref{tab:specflux} it was possible to obtain for ISGRI values of the photon 
index 
for the time intervals listed below:

\begin{minipage}[t]{8.8cm}
\hspace{-0.8cm}
\begin{tabular}{rll}
Position: & Interval (UTC) & PhoIndex \\
\vspace{1mm}
decay of the $1^{\rm st}$ peak: & 10:12:06-10:12:16 &  1.859 ${0.10 \atop -0.10}$ \\
\vspace{1mm}
small peak in between: & 10:12:16-10:12:26  & 2.329${0.18 \atop -0.23}$ \\
\vspace{1mm}
rising of the  $2^{\rm nd}$  peak: & 10:12:26-10:12:31 & 1.955${0.28 \atop -0.25}$ \\
\vspace{1mm}
decay of the  $2^{\rm nd}$  peak: & 10:12:37-10:12:52  & 1.849${0.22 \atop -0.22}$ \\
\end{tabular}
\end{minipage}
\newline

With these ISGRI data it was possible to track the spectral evolution over the whole emission period,
showing a general hard-to-soft
evolution for the first peak up to the small peak between the two main peaks (Fig.\,\ref{fig:spi-ibis-lc} lower panel). 
During the second
peak the spectrum hardens again (yet being softer compared to the first peak), but showing
no evolution. The obtained ISGRI fluxes are reasonably correlated with the light curve. 

The photon indices obtained with XSPEC 12.0 for SPI agree within the errors with
the results obtained with ISGRI for interval 2, 3, 5 and 6, 
but reveals no
spectral evolution. In contrast to the time between interval 2 and 3 (between to two main peaks) and after the last interval 3 
(decay of the $2^{\rm nd}$ peak), it was possible to extract for SPI a spectrum for interval 1, 
although SPI's lightcurve did not show a rate increase. As expected the obtained photon index 
and flux 
do not agree well with the ISGRI values. A large discrepancy is observed for interval 4 (maximum of the  $1^{\rm st}$ peak). 
SPI 
shows here a rather hard spectrum whereas for ISGRI the spectrum is even softer compared to interval 5.
The SPI results have to be handled with care because for such a short 2\,s interval, containing only a small number of counts 
in each energy bin, the determination of a spectrum  is more difficult.

Comparing the results obtained with the three different methods used for the SPI data analysis, one can see that the XSPEC 
12.0 derived 
photon indices yield the best agreement with the ISGRI data, although the same is valid for the SPIROS+XSPEC 11.2 results, 
using a diagonal 
response (line SPI$_{11.2}^{\rm diag.}$ of Tab. ~\ref{tab:specflux}), but these have a much broader confidence range. 
Second it should be noted that XSPEC 12.0 is using an 
off-diagonal response, which should represent the instrument much better.

The burst had a 20--200\,keV peak flux (over 2\,s) of  
$4.1 {1.3\atop -1.4}\cdot 10^{-7}$\,erg\,cm$^{-2}$\,s$^{-1}$
and $5.4 \cdot 10^{-7}$\,erg\,cm$^{-2}$\,s$^{-1}$
and a 25--100\,keV peak flux of 
$1.7 {0.9\atop -0.7} \cdot 10^{-7}$\,erg\,cm$^{-2}$\,s$^{-1}$
and $3.0 \cdot 10^{-7}$\,erg\,cm$^{-2}$\,s$^{-1}$ measured
with SPI and IBIS/ISGRI, respectively. These peak fluxes would place
GRB030320 in the top 25\% of the BATSE peak flux distribution
(\cite{paciesas99}).

The gamma-ray fluence in the 20--200\,keV band measured with SPI is
$1.35{0.21\atop -0.26} \cdot 10^{-5}$\,erg\,cm$^{-2}$, consistent with 1.1$\cdot$10$^{-5}$\,erg
cm$^{-2}$ from IBIS/ISGRI. The 25--100\,keV fluence was $7.1{1.6\atop -1.8}\cdot 10^{-6}$\,erg cm$^{-2}$ (SPI) 
and $6.5 \cdot 10^{-6}$\,erg cm$^{-2}$ (ISGRI). Peak flux and
fluence agree within a factor of two with the results obtained with
Ulysses (\cite{gcn1943}).

\section{Conclusion}

The analysis of GRB030320 showed that INTEGRAL is still able to detect GRBs at
the edge of its FoV. For this GRB a photon index of $\Gamma \simeq 1.7$ was derived for the prompt emission.
The time resolved spectroscopy revealed a hard-to-soft transition of the spectrum during the 60\,
burst duration.
Especially IBIS/ISGRI showed a good performance in localisation
and in the determination of the spectral evolution. The difficulties observed with SPI
will hopefully improve with a better understanding of the
response at the edge of the FoV.

\begin{acknowledgements}
The SPI project has been completed under 
the responsibility and leadership of CNES. We are grateful 
to ASI, CEA, CNES, DLR, ESA, INTA, NASA and OSTC for support.
The SPI/ACS project is supported by the German "Ministerium f\"ur Bildung und Forschung" through 
DLR grant 50.OG.9503.0.

\end{acknowledgements}


\bibliographystyle{amsplain}


\end{document}